\documentclass{article}
\pdfoutput=1
\usepackage{jinstpub}
\usepackage{graphicx}

\newcommand{\totalChips}{17200}
\newcommand{\totalA}{4389}
\newcommand{\totalB}{8449}
\newcommand{\totalC}{192}
\newcommand{\totalD}{641}

\newcommand{\totalFailed}{990}

\newcommand{\totalFunctional}{14661}
\newcommand{\totalPlotsABCD}{13666}
\newcommand{\totalAB}{12838}
\newcommand{\totalSpikesFraction}{3} 

\author[1,2]{T.~Andeen}
\author[1]{J.~Ban}
\author[1]{G.~Brooijmans}
\author[1]{A.~Emerman}
\author[3]{P.~Kinget}
\author[3]{J.~Kuppambatti}
\author[1]{D.~Mahon}
\author[1]{I.~Ochoa}
\author[1]{W.~Sippach}
\author[1]{Q.~Wang}

\affiliation[1]{Nevis Laboratories, Columbia University}
\affiliation[2]{University of Texas at Austin}
\affiliation[3]{Dept. of Electrical Engineering, Columbia University}

\emailAdd{miochoa@nevis.columbia.edu}

\keywords{Calorimeters; Front-end electronics for detector readout; Radiation-hard electronics}
\title{Performance and Quality Control of a Radiation-Hard 12-bit 40~MSPS ADC for the ATLAS Liquid Argon Calorimeter Trigger Readout Electronics Phase-I Upgrade at the LHC}

\abstract{A radiation-hard quad-channel 12-bit 40~MSPS pipeline analog-to-digital converter (ADC) has been designed for the trigger readout electronics Phase-I upgrade of the ATLAS Liquid Argon calorimeter, at the CERN Large Hadron Collider. The final version of the custom design, fabricated in a commercial 130~nm CMOS process, is presented and found to meet the system requirements for analog performance and radiation tolerance. The procedure for quality control of \totalChips~ADCs is described and the results are presented.}

\begin{document}

\maketitle

\section{Introduction} \label{sec:Intro}

This article describes the design and performance of a radiation-hard quad-channel 40~MSPS pipeline analog-to-digital converter (ADC), used in the new trigger readout electronics of the ATLAS Liquid Argon (LAr) calorimeter installed as part of the ATLAS Phase-I upgrade \cite{phase1tdr}.

The Large Hadron Collider (LHC) \cite{Evans:2008zzb} at CERN is a proton-proton collider that has operated at centre-of-mass energies of 7, 8 TeV from 2010 to 2012 (Run 1) and 13 TeV from 2015 to 2018 (Run 2), with a proton bunch crossing frequency of 40 MHz. During Run 1 and Run 2, the LHC has delivered collisions at increasing rates, with a peak instantaneous luminosity of $2\times10^{34}$~cm\textsuperscript{-2} s\textsuperscript{-1} achieved in 2018, exceeding the design luminosity by a factor two. ATLAS \cite{Collaboration_2008} is a general purpose experiment composed of several subsystems that can collect data from particles produced by the high energy collisions. The ATLAS LAr calorimeter is a high-granularity sampling calorimeter that measures the energy and position of electrons, photons and jets of hadrons as they travel through the detector. The readout of the signals is performed by on-detector electronics \cite{FEB} that receive, amplify and shape the calorimeter cell signals. The waveform is sampled at 40 Mega samples per second (MSPS). The information is stored in an analog buffer and then digitized on reception of a trigger accept signal, issued by the Level 1 (L1) trigger system, based on the magnitude of the signal in large regions of the detector, referred to as trigger towers. Signals for trigger towers are built by summing those from individual cells in the on-detector electronics; this part of the readout is upgraded as part of the Phase-I upgrade, and uses the ADC discussed in this article.

\begin{figure}[htbp!]
\begin{center}
\includegraphics[width=1\textwidth]{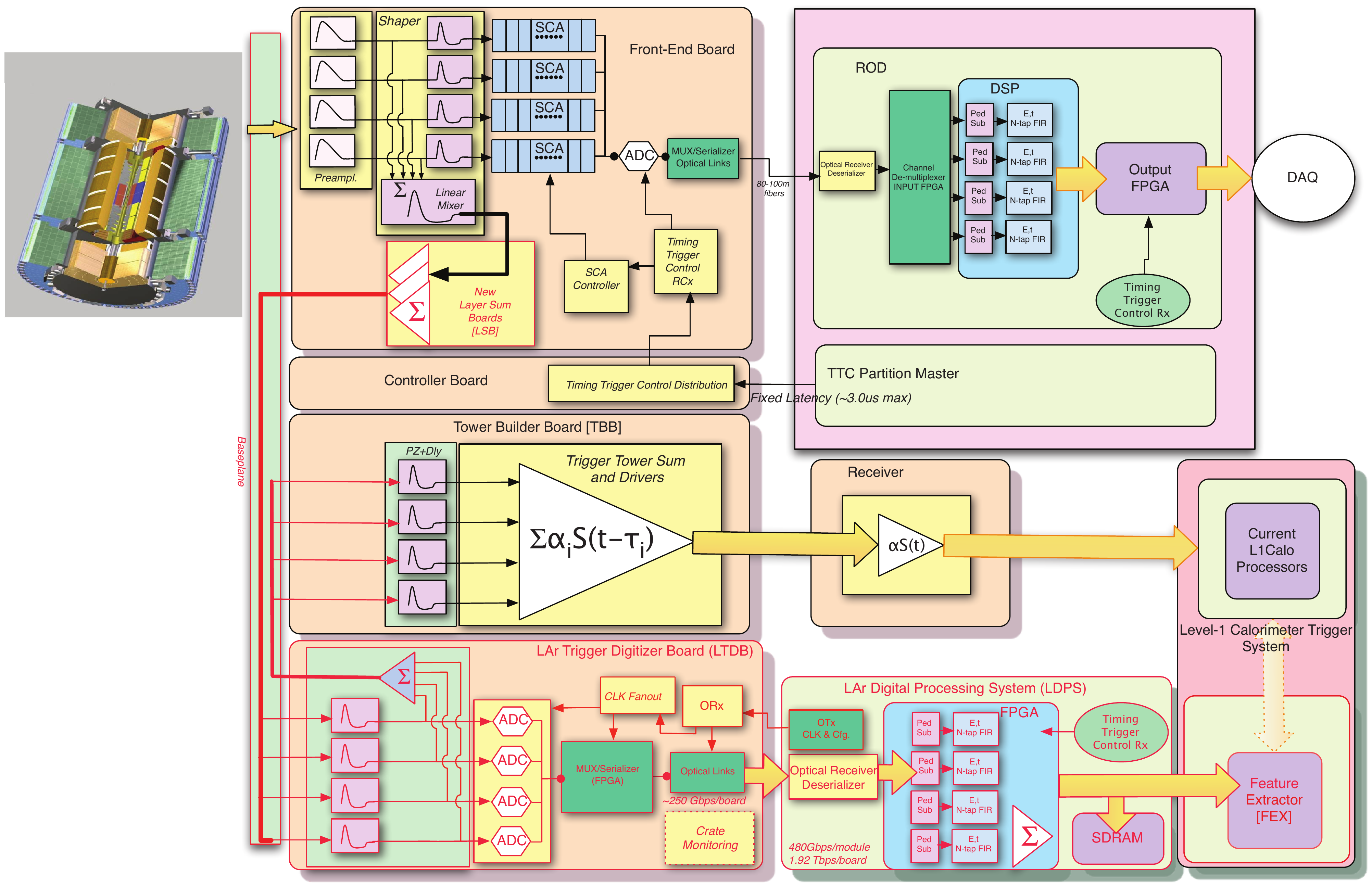}
\end{center}
\caption{Schematic diagram of the LAr trigger and readout architectures for the Phase-I upgrade of the LAr calorimeter \cite{phase1tdr}. The new components are highlighted in red.}
\label{fig:upgrade_diagram}
\end{figure}

The Phase-I upgrade of the accelerator and detectors is taking place during a long shutdown in 2019 and 2020, with Run 3 set to begin in 2021. The LHC Phase-I upgrade will allow for instantaneous luminosities reaching $2.5\times10^{34}$~cm\textsuperscript{-2} s\textsuperscript{-1}. In order to cope with the increased collision rate with a constant trigger bandwidth, the trigger input signals from the LAr calorimeter are improved. The granularity of the calorimeter signals sent to L1 is increased with respect to the trigger towers, allowing for better background rejection. These ``super cell" signals are digitized at MSPS on new on-detector LAr Trigger Digitizer Boards (LTDBs) and all the corresponding data are sent off-detector for digital filtering. On each LTDB, 80 ADCs are responsible for the 12-bit digitization of the analog signals from 320 supercells at 40~MSPS. The digital signals are then transmitted with custom ASICs \cite{locid} using 40 optical links at 5.12~Gbps each. The LTDB and off-detector systems responsible for the processing of the digitized signals are shown in Figure~\ref{fig:upgrade_diagram}.

A Phase-II upgrade will take place in a later shutdown, after Run 3 and ahead of the High-Luminosity LHC (HL-LHC) run. After this second upgrade, the LHC will be able to increase the instantaneous luminosity by another factor of 2 to 4. The Phase-II upgrade of the LAr calorimeter readout will have all cells digitized at 40~MSPS, moving the pipeline off-detector.
The Phase-I upgrades are fully compatible with the plans for Phase-II.

The work described in this article is the culmination of the prototype designs described in refs. \cite{nevis10} and \cite{nevis12}. The final design stage and the production chip performance are presented, as well as quality control tests and radiation tolerance studies.

\section{System Specifications}

The ADC is designed to receive the shaped analog signals from the LAr calorimeter super cells, which must be continuously sampled and digitized at a frequency of 40~MSPS. Due to strict power and latency budgets, each ADC must consume less than 145 mW and have latency less than 200 ns. In order to cover the energy range of interest to the trigger system with sufficient precision, a dynamic range of approximately 12 bits is required together with 10-bit precision \cite{phase1tdr}.

The ADC is expected to operate reliably through the rest of the LHC runs, for a total expected integrated luminosity of 3000~fb\textsuperscript{-1}. Radiation tolerance is one of the requirements for any electronics component located on-detector, placing strict demands on the tolerance of the component to total ionizing dose and single-event effects (SEE), which depend on detector location and corresponding levels of radiation \cite{radiation_qualification}. LAr on-detector ASICs are required to be radiation tolerant to a total ionizing dose of 180~kRad, non-ionising energy loss doses of up to $4.9\times 10^{12}$~n\textsubscript{eq} per cm\textsuperscript{2}, and suffer at most a small number of recoverable SEEs for a total fluence of $7.7 \times 10^{12}$~hadrons per cm\textsuperscript{2} \cite{phase2tdr}.

\section{ADC Implementation and Operation}

\begin{figure}[htbp!]
\begin{center}
\includegraphics[width=1\textwidth]{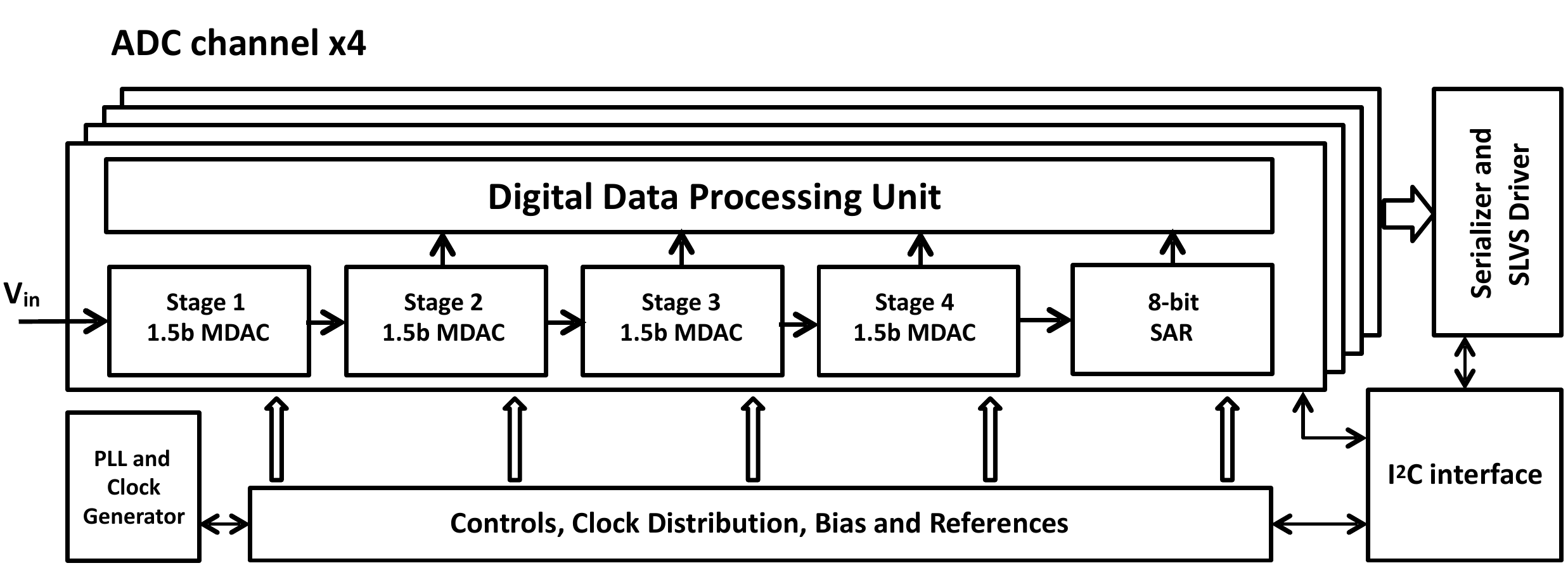}
\end{center}
\caption{ADC block diagram.}
\label{fig:diagram_channel}
\end{figure}

\begin{figure}[htbp!]
\begin{center}
\includegraphics[width=0.45\textwidth]{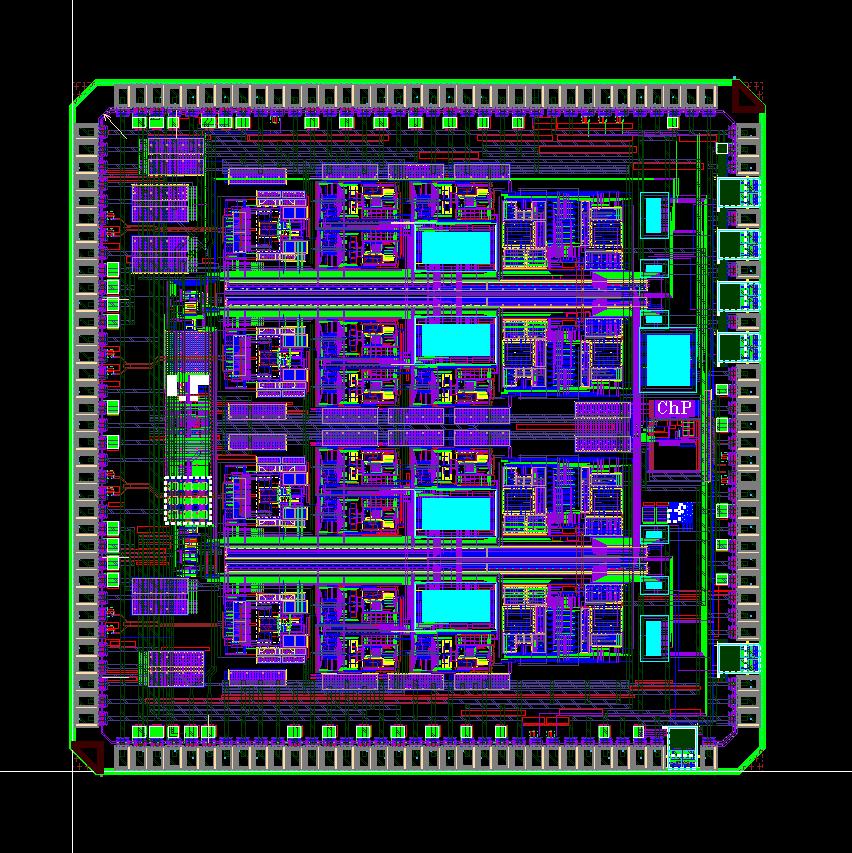}
\end{center}
\caption{Image of the ADC layout (3.6~mm x 3.6~mm).}
\label{fig:chip_photo}
\end{figure}

The chip consists of four identical ADC channels, each composed of a multiplying digital-to-analog converter (MDAC) pipeline followed by a successive approximation (SAR) ADC \cite{nevis12}, as shown in Figure~\ref{fig:diagram_channel}, implemented in Global Foundries CMOS 8RF 130 nm technology. The pipeline consists of four 1.5-bit stages with a nominal gain of two each, resolving the 4 most significant bits \cite{nevis12}. The remaining 8 least significant bits are resolved by the SAR stage. An internal Phase Lock Loop (PLL) generates a 640~MHz clock from a 40~MHz input clock, both edges of which are used for the SAR operation. The MDAC stages operate with a supply voltage of 2.5~V and a common-mode of 1.25~V, while for the SAR a supply voltage of 1.2~V with a common-mode of 0.6~V is used. The digital output of the ADC is formed in a digital data processing unit (DDPU) and the signals are serialized for output over Scalable Low-Voltage Signaling (SLVS) at 320~MHz. An I\textsuperscript{2}C interface is used for chip control. Figure~\ref{fig:chip_photo} shows an image of the chip layout.

A digital calibration is performed to correct for capacitor mismatch in the MDAC stages. The algorithm measures the actual gain and compares it to an ideal gain of two, with the difference taken as a correction to the digital output. By design, the gains are lower than two, since gains larger than two would lead to missing codes. As a result, the dynamic range of the chip is slightly reduced with respect to the nominal 12 bits (4096 counts). The calibration constants are calculated offline but stored on chip and applied in the DDPU.

\begin{figure}[htbp!]
\begin{center}
\includegraphics[width=0.65\textwidth]{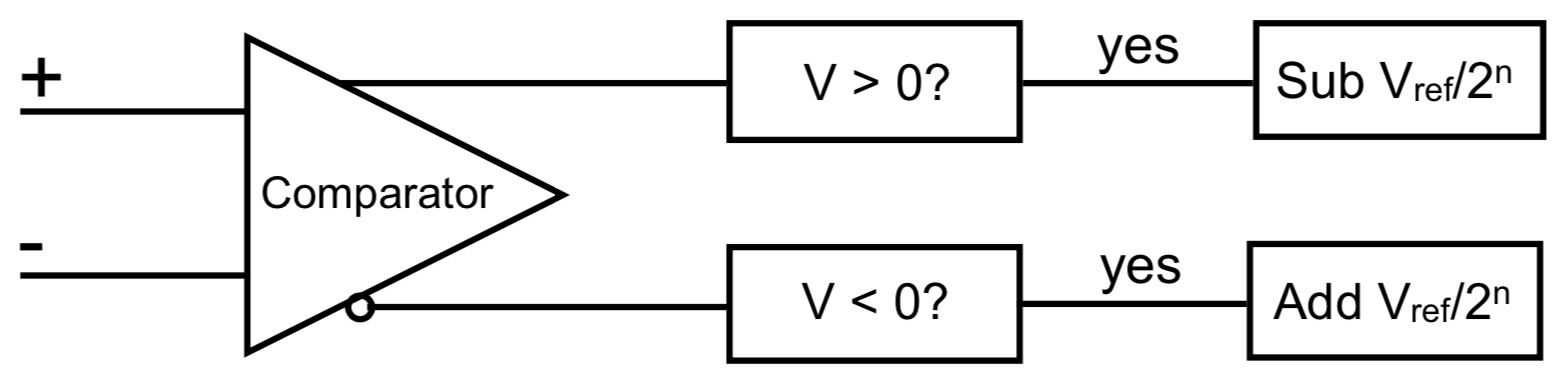}
\end{center}
\caption{Schematic diagram of the SAR comparator.}
\label{fig:SARcomp}
\end{figure}

SAR comparator metastability, observed in a small fraction of prototype chips, is managed by a special circuit. When the SAR comparator input is very close to zero, its output may not have settled when strobed, returning ``yes" for both conditions (see Figure~\ref{fig:SARcomp}), resulting in output code jumps. In prototype chips, the rate of code jumps could be driven to tens of Hz by tuning the input signal. The comparator metastability diagnosis was unambiguously confirmed by observing shorts on the ADC $V_{ref}$ output simultaneous with code jumps. To remove this problem, the final version of the chip includes an additional step in the SAR logic that disallows simultaneous addition and subtraction.

\begin{figure}[htbp!]
\begin{center}
\includegraphics[width=0.55\textwidth]{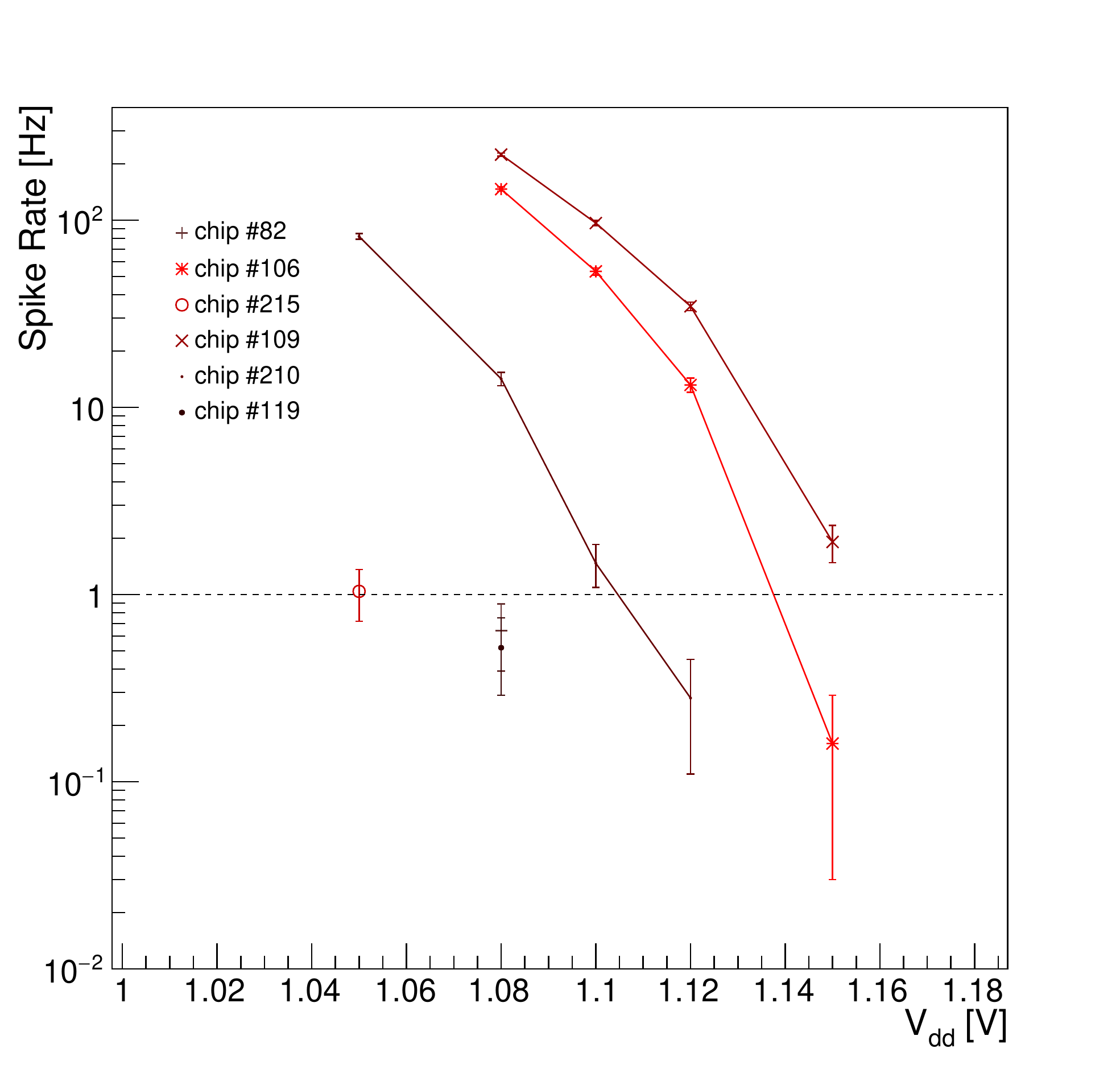}
\end{center}
\caption{Rate of spikes measured at several $V_{dd}$ values, for single channels of several ADC chips.}
\label{fig:VddScan}
\end{figure}

Output code errors (``spikes") of a different nature are observed in a small percentage of the chips. These occur at no particular SAR input voltage and at different ADC outputs, with rates of few Hz. It was observed that decreasing the digital logic speed, which can be effected by reducing the digital supply voltage, $V_{dd}$, below its nominal 1.2 V value, led to higher spike rates. Chips not showing any spikes during normal operation also started showing spikes for $V_{dd}<1.2$~V, again with increasing rates for decreasing voltages. An example of the $V_{dd}$ dependence of the rate is shown in Figure~\ref{fig:VddScan} for different chips. The highest $V_{dd}$ at which the rate drops below 1 Hz, $V_{th}$, was found to be 1.15 V. 

The common-mode conversion between the different voltage domains of the pipeline (2.5 V) and the SAR (1.2 V) is understood to be a possible cause for the spikes. 
Increasing the SAR analog supply voltage  $V_{a}$ to 1.3~V increases the margin for the common-mode voltage conversion. Under that condition, spikes are no longer observed in chips previously found to have them. The nominal operation point is therefore set at $V_{a}=1.3$~V, while keeping the logic supply at $V_{dd}=1.2$~V, leading to per-cent level additional power dissipation. The quality control procedure, discussed in Section \ref{sec:qc}, includes a measurement of each chip's margin with respect to this issue. 

\section{Radiation Tolerance}

The radiation tolerance of the ADC chip has been studied in its different prototyping stages. The TID and SEE tests performed with previous prototypes are described in refs.~\cite{nevis10} and \cite{nevis12}. 

\begin{figure}[htbp!]
\begin{center}
\includegraphics[width=0.5\textwidth]{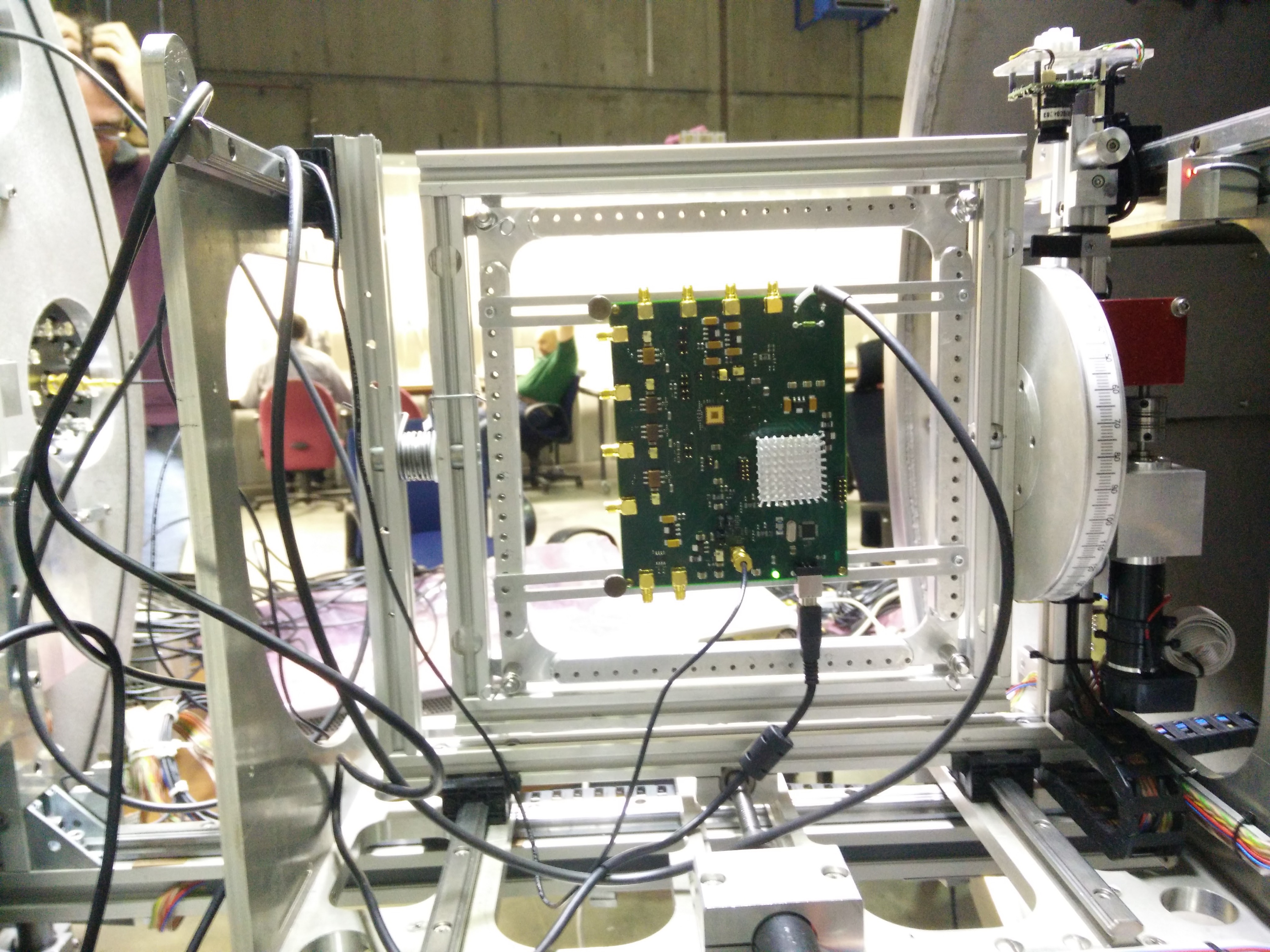}
\end{center}
\caption{Test board for irradiation.}
\label{fig:board}
\end{figure}

In addition to proton radiation tests, SEE cross-sections were measured using heavy ions, targeting a larger range of deposited energies and complementing the existing studies with protons. One chip packaged with a removable lid was taken to the Cyclotron Resource Centre at Louvain-la-Neuve, Belgium. It was mounted on a specialized board (Figure~\ref{fig:board}). The board was then placed inside a vacuum chamber and positioned such that the beam of heavy ions was centered on the chip and incident at a perpendicular angle. Four different heavy ions were used, with characteristics detailed in Table \ref{tab:hi}. The beam flux ranged from 2 to 15$\times10^{3}$~particles/cm$^{2}$s, with a 25 mm diameter.
 
\begin{table}[!htb]
\begin{center}
\caption{High energy heavy ions used for the SEE tests and corresponding energy on device under test (DUT), range and linear energy transfer (LET) in silicon.}
\begin{tabular}{|cccc|}
\hline
Ion & Energy on DUT & Range Si & LET Si \\
& [MeV] & [$\mu$m] & [MeV/(mg/cm\textsuperscript{2})] \\
\hline
\textsuperscript{14}N\textsuperscript{4+} & 122 & 171 & 1.9 \\
\textsuperscript{22}Ne\textsuperscript{7+} & 238 & 202 & 3.3 \\
\textsuperscript{58}Ni\textsuperscript{18+} & 582 & 101 & 20.4 \\
\textsuperscript{124}Xe\textsuperscript{35+} & 995 & 73.1 & 62.5 \\
\hline
\end{tabular}
\label{tab:hi}
\end{center}
\end{table}

During the test, the chip was powered and clocked at 40 MHz and the current draw monitored. A firmware filter identified single-event-effects as ADC output codes outside a window of $\pm 3$ counts around the expected value (the ADC inputs were left open yielding a fixed output code in the middle of the ADC range). No latchup events were observed. For the Nickel run, the number and type of events observed on each channel of the chip are shown in Table~\ref{tab:run5}, as well as the calculated cross-sections. In the Table, ``analog" SEUs are distinguished from ``digital" SEUs: ``analog" SEUs are upsets in the least significant bits of the ADC, and their rate is inversely proportional to the size of the capacitors used to measure those bits, i.e. less significant bits have higher upset rates.  These SEUs effectively degrade the analog performance of the ADC, and are therefore tagged as ``analog".  ``Digital" SEUs occur with equal probability for any ADC output bit. A signal-event-functional-interrupt (SEFI) is observed when the ADC output gets stuck at a given value, and a reset is needed to return to normal
operation (but not a power cycling of the device).

\begin{figure}[htbp!]
\begin{center}
\includegraphics[width=0.47\textwidth]{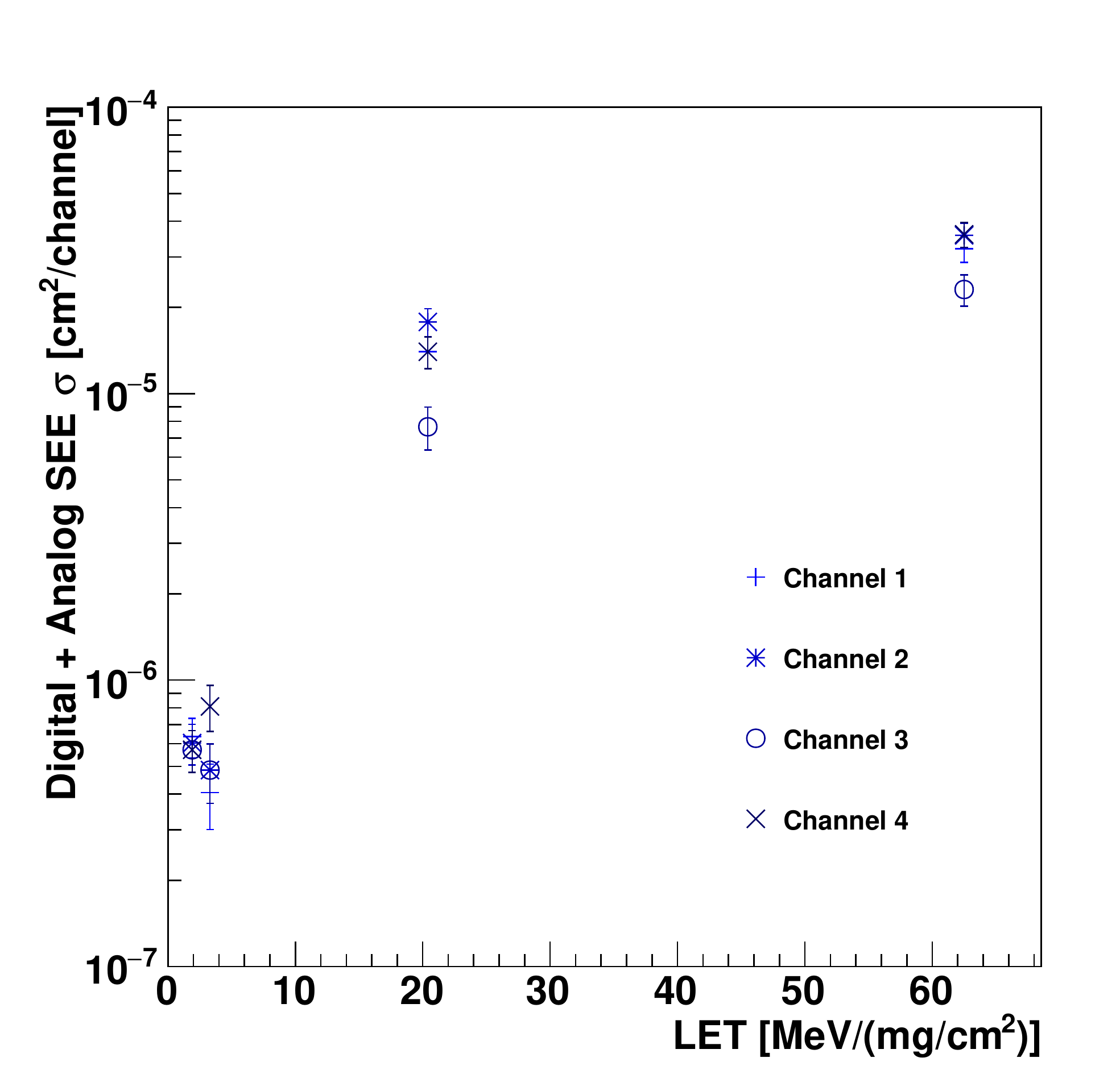}
\includegraphics[width=0.47\textwidth]{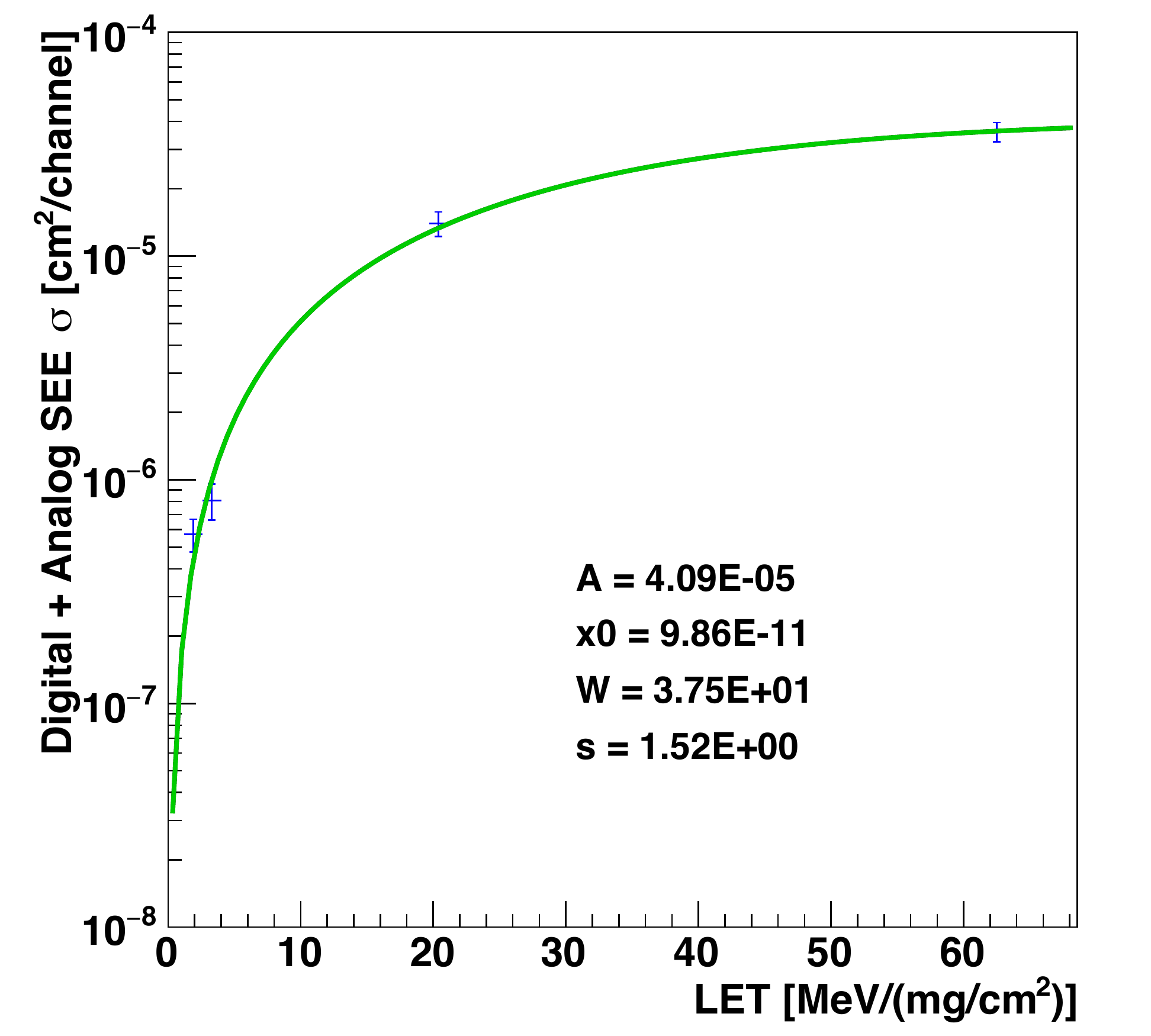}
\end{center}
\caption{Total SEE cross-sections for each ADC channel, as a function of LET.}
\label{fig:LETall}
\end{figure}

The dependence of the SEE rate with the LET of the heavy ion can be parameterized with a Weibull function \cite{WeibullPaper}:
\begin{equation}
\label{eq:weibull}
F(x)=A[1-\exp(-(\frac{x-x_{0}}{W})^{s})]
\end{equation} where $A$ is the plateau cross-section, $x_{0}$ is such that $F(x)=0$ for $x<x_{0}$, $W$ is a width parameter and $s$ is a dimensionless exponent. The total cross-section values measured for all four channels are shown in Figure~\ref{fig:LETall} (left), as a function of the linear energy transfer (or heavy ion). A fit to the results for one of the ADC channels is shown in Figure~\ref{fig:LETall} (right), including the fit parameter results. The proton SEE cross-section can be estimated by extrapolating the fit results to the electronic dE/dx of 200~MeV protons in Si, resulting in cross-section values that are compatible with the $10^{-12}$~cm$^{2}$ proton irradiation measurements\footnote{The electronic stopping power of 200~MeV protons in silicon is 3.63 MeV~cm$^{2}$~/~g \cite{dEdx}.} \cite{nevis12}.

\begin{table}[!htb]
\begin{center}
\caption{Number of SEEs observed per ADC channel in a chip under a \textsuperscript{58}Ni\textsuperscript{18+} beam, broken down in terms of analog or digital SEUs and SEFIs, as defined in the text. The cross-section values take into account all error types.}
\begin{tabular}{|cccccc|}
\hline
\textsuperscript{58}Ni\textsuperscript{18+}& SEU & SEU & SEFI & SEE & Cross-section (w/ analog errors) \\
& (Analog) & (Digital) & & & [cm$^{2}$] \\ 
\hline
Channel 1 & 59 	& 3  & 1 & 63 & $6.76\times10^{-7}$ ($1.40\times10^{-5}$) \\
Channel 2 & 75 	& 4  & 1 & 80 & $9.01\times10^{-7}$ ($1.78\times10^{-5}$) \\
Channel 3 & 32 	& 2  & 1 & 35 & $4.50\times10^{-7}$ ($7.66\times10^{-6}$) \\
Channel 4 & 61 	& 1  & 1 & 63  & $2.25\times10^{-7}$  ($1.40\times10^{-5}$) \\
\hline
\end{tabular}
\label{tab:run5}
\end{center}
\end{table}

\begin{figure}[htbp!]
\begin{center}
\includegraphics[width=0.5\textwidth]{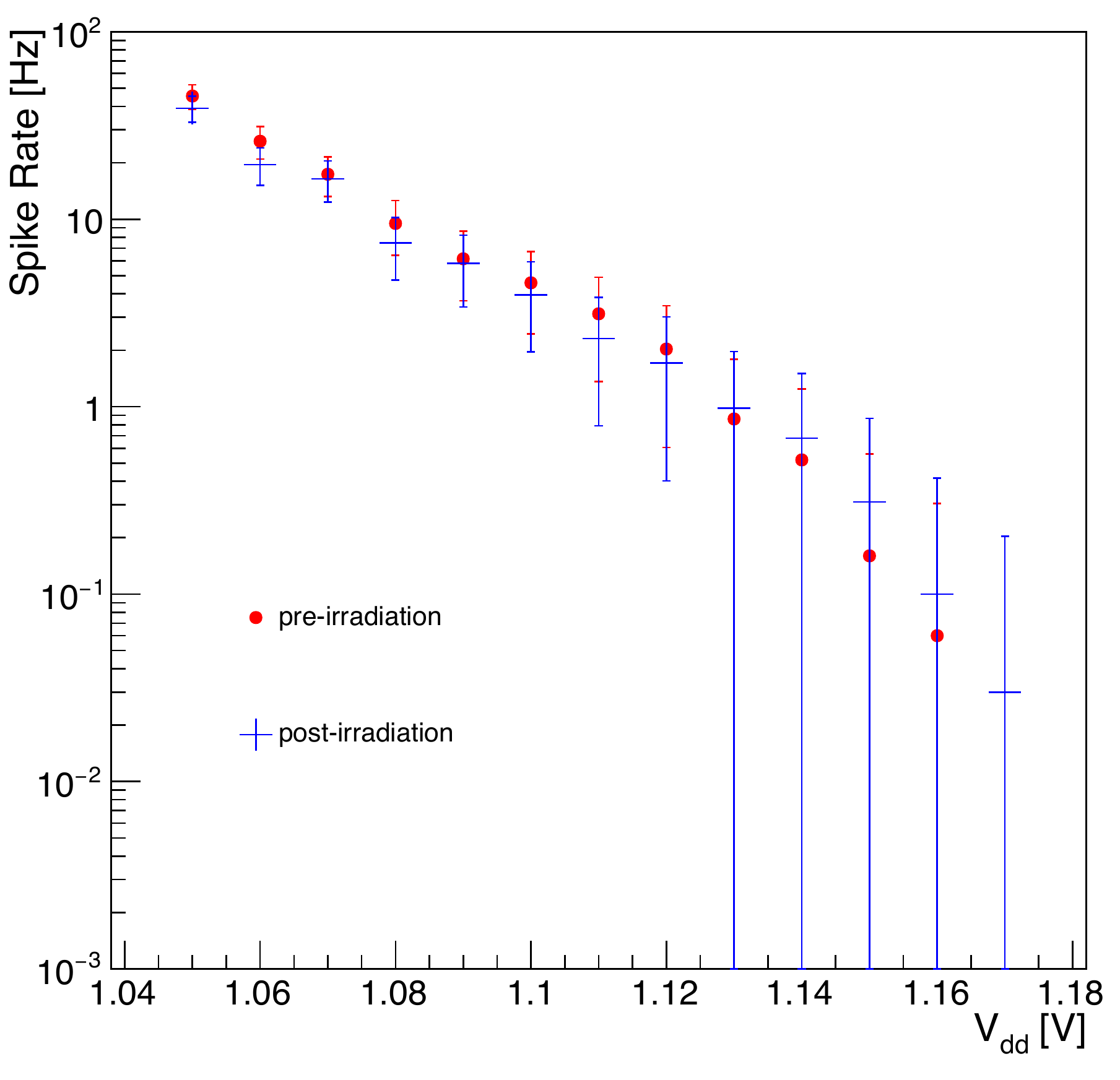}
\end{center}
\caption{Rates of spikes on the ADC output as a function of the digital voltage supplied, measured before and after a total ionizing dose of 250 kRad was delivered.}
\label{fig:preVspost}
\end{figure}

A proton irradiation run was also performed to characterize the low $V_{dd}$ spike behavior's possible dependence on fluence. One chip was irradiated at Massachussetts General Hospital with 200 MeV protons, with a total dose of 250 kRad delivered to the chip. The spike rates were measured for a range of $V_{dd}$ values, both before and after the irradiation, for a single channel (as shown in Figure~\ref{fig:preVspost}). A dose of 250 kRad was shown to have negligible impact on the spike rates and on $V_{th}$.

\section{Quality Control} \label{sec:qc}

A total of 124 LAr Trigger Digitizer Boards (LTDB) are needed for the LAr calorimeter ATLAS Phase-I upgrade, each with 320 channels, using 80 quad-channel ADCs. To assemble the necessary LTDBs for the detector and spares, approximately 12000 good chips are needed.  To allow for yield factors, about 17000 chips were produced and packaged in molded QFN-72 packages.

The quality control (QC) protocol aims to determine which chips are fully operational and assess their analog performance. The expected types of failure modes had been established in previous smaller-scale tests, as well as the typical chip-to-chip variability. Other characteristics, such as power consumption and latency, have been found to meet the requirements during the design phase, and are not targeted by this QC protocol.

\begin{figure}[htbp!]
\begin{center}
\includegraphics[width=0.5\textwidth]{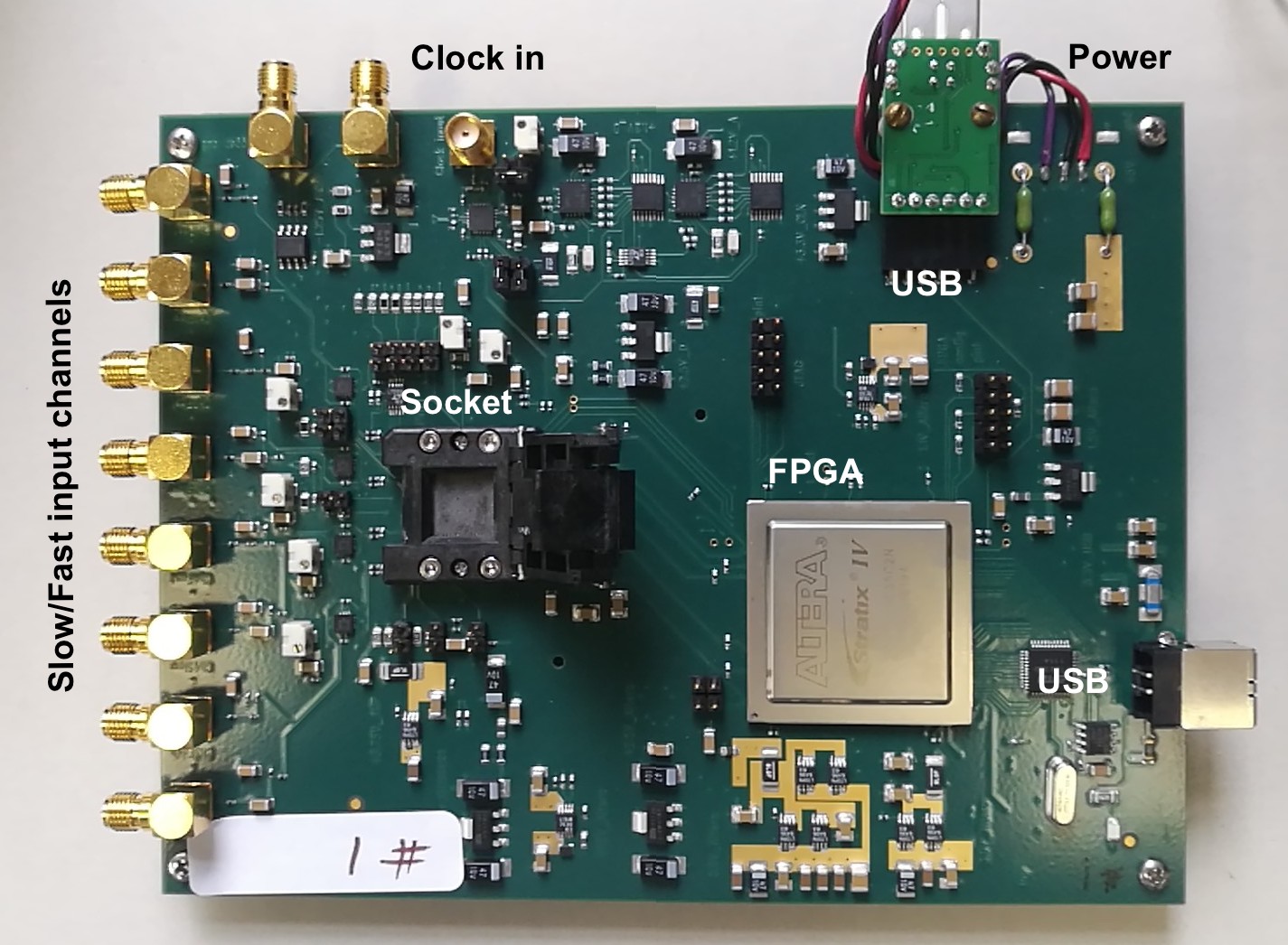}
\end{center}
\caption{Photo of a socketed board used for testing.}
\label{fig:testSetup}
\end{figure}

Individual chips are subjected to a series of tests while in a socket on a test board, shown in Figure~\ref{fig:testSetup}.  The test board allows injection of sine waves of various frequencies and amplitudes, injection of a low-jitter clock signal, and chip read-out via an FPGA. The full procedure for each chip is as follows, with a maximum duration of 2 minutes per chip:
\begin{enumerate}
\item The chip is placed in the socket and the I\textsuperscript{2}C connection is established after a chip and software reset.
\item The MDAC calibration is performed for each channel.
\item A fast (approximately 5 MHz) sine-wave is sent to the ADC inputs, scanning several amplitude values.
\item A slow (200 Hz) sine-wave is sent to the ADC inputs, while a firmware filter identifies and counts any large jumps in the ADC output during an interval of 20 s. This is performed for three $V_{dd}$ values (1.14, 1.17, 1.20 V) and for each channel a spike rate is measured.
\end{enumerate}

\begin{figure}[htbp!]
\begin{center}
\includegraphics[width=0.62\textwidth]{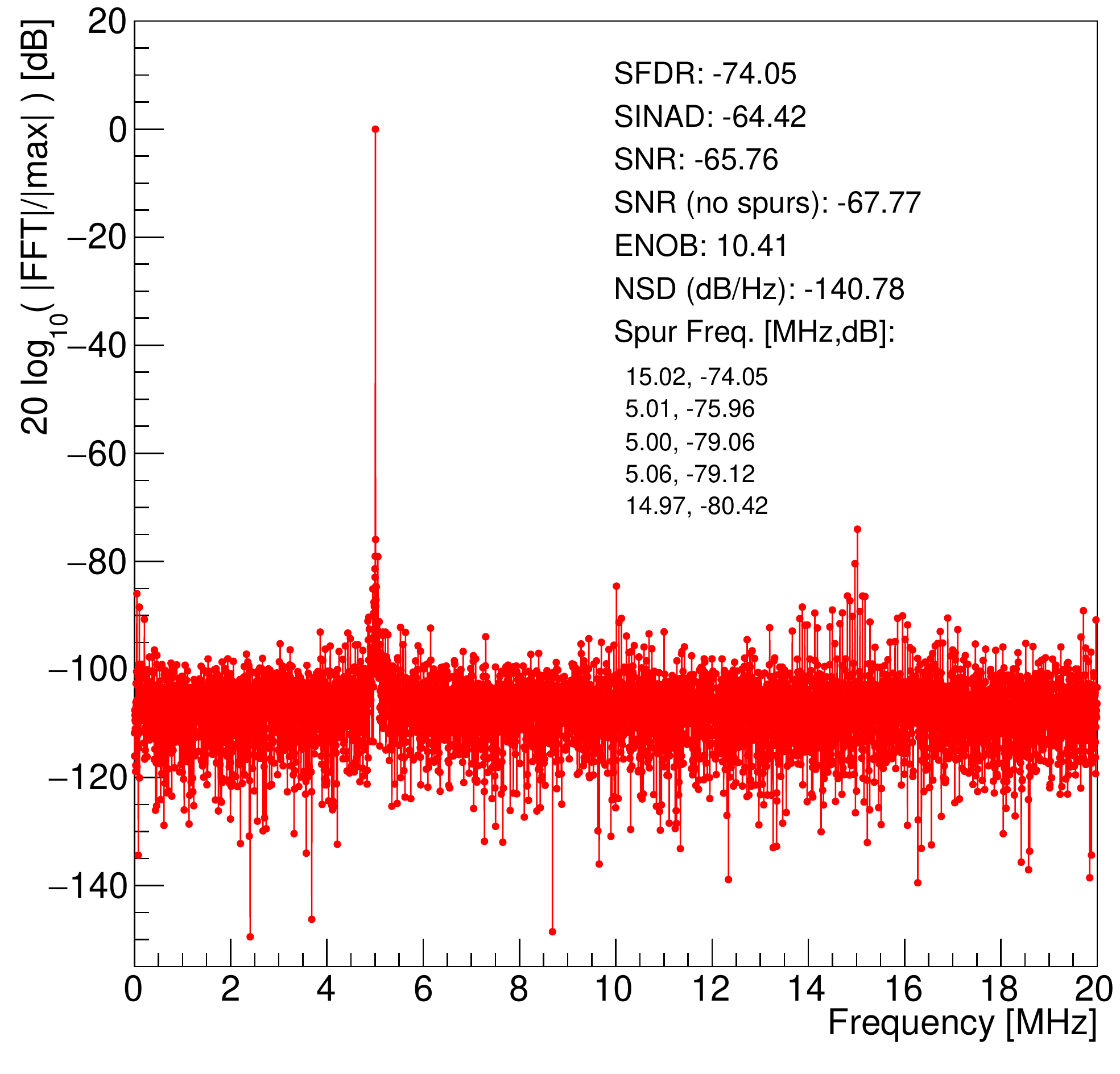}
\end{center}
\caption{FFT example.}
\label{fig:FFT}
\end{figure}

In step 3, the analog chip performance is measured by digitizing a pure sine-wave and transforming the output into the frequency domain, via a Fast Fourier Transform, using 12288 samples, as illustrated in Figure~\ref{fig:FFT}. In frequency domain, the signal component is compared to the harmonics and noise contributions, from which several figures of merit can be calculated, as described in Section \ref{sec:measur}. As part of the quality assurance (QA) process, prototypes and a small fraction of production chips were tested with carrier frequencies ranging from 200 kHz to 18 MHz. For the purpose of the QC, only a 5 MHz carrier frequency is used, for its proximity to the shaped pulse expected at the ADC inputs in the LAr calorimeter.

As a consequence of the observation of spikes during the QA process, the QC protocol includes the digitization of a slow (200 Hz) sine-wave, monitoring the output for jumps larger than 7 counts between two consecutive ADC samples. 
This measurement is performed for a set of $V_{dd}$ values: 1.14, 1.17, 1.20 V. 

\section{Results} \label{sec:measur}

A total of \totalChips~chips were tested using two test benches, requiring a total of 6 sockets for the full QC (each socket being rated for 50000 insertions). Of those, approximately 15\% were non-functional chips, a yield loss dominated by packaging issues (based on the error patterns observed), and consistent with yield factors found during the QA process. Only the \totalFunctional~functional chips were consider further.

Each chip was classified according to the dynamic range (DR) and effective number of bits (ENOB) of its worst performing channel. Table \ref{tab:cat} lists the categories and their definitions, as well as how many chips fall in each. Only chips in categories A and B are accepted as having successfully passed the QC tests and are used in LTDB assembly, corresponding to a yield of approximately 88\%. Chips of C and D categories are still functioning chips and can therefore be useful for spares. Chips categorized as F are those that show errors during any step of the QC test, including those with at least one channel showing non-zero spike rates for any $V_{dd}$ value, which occurs in less than \totalSpikesFraction~\% of the chips. 

\begin{figure}[htbp!]
\begin{center}
\includegraphics[width=0.48\textwidth]{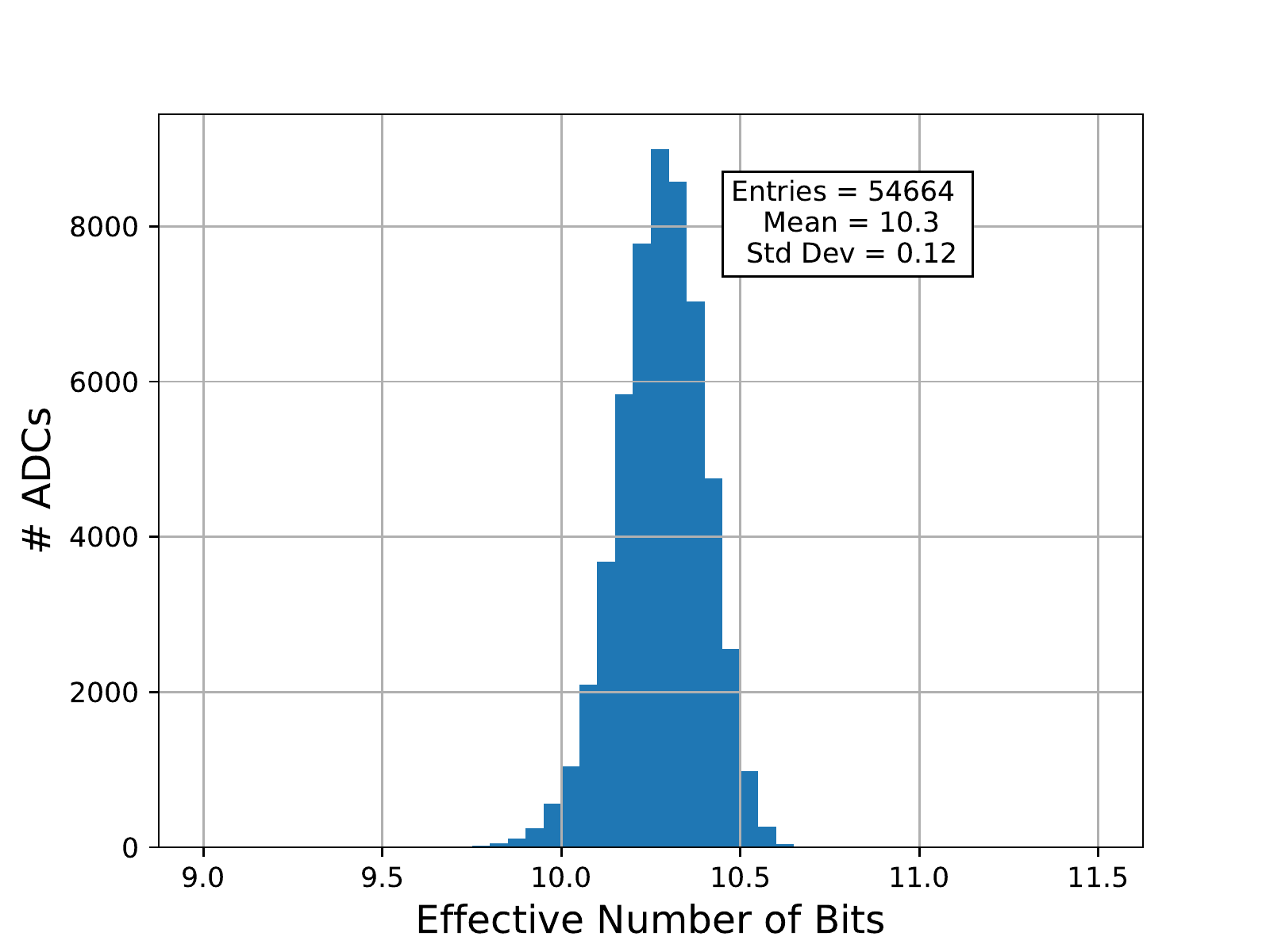}
\includegraphics[width=0.48\textwidth]{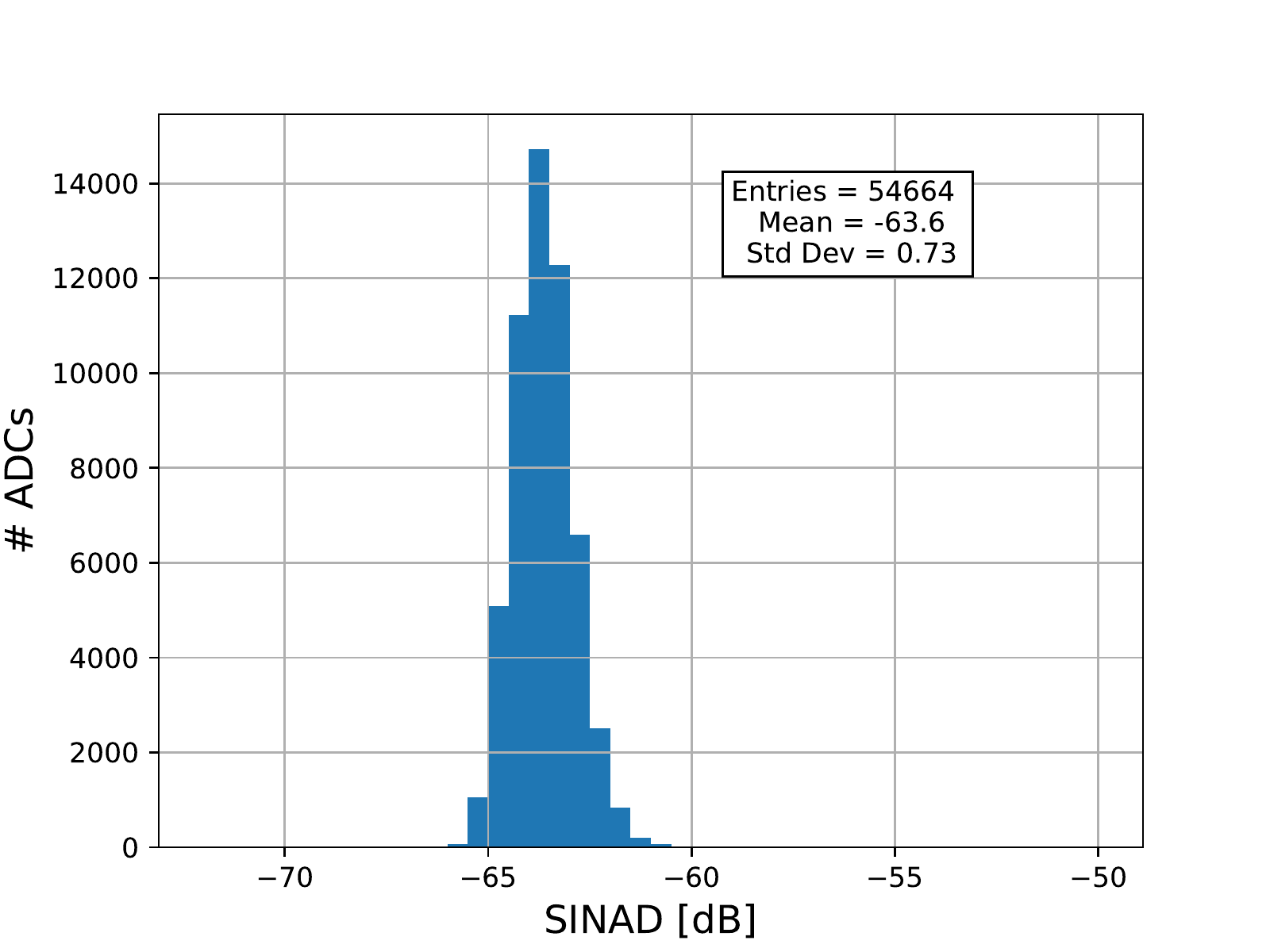}
\includegraphics[width=0.48\textwidth]{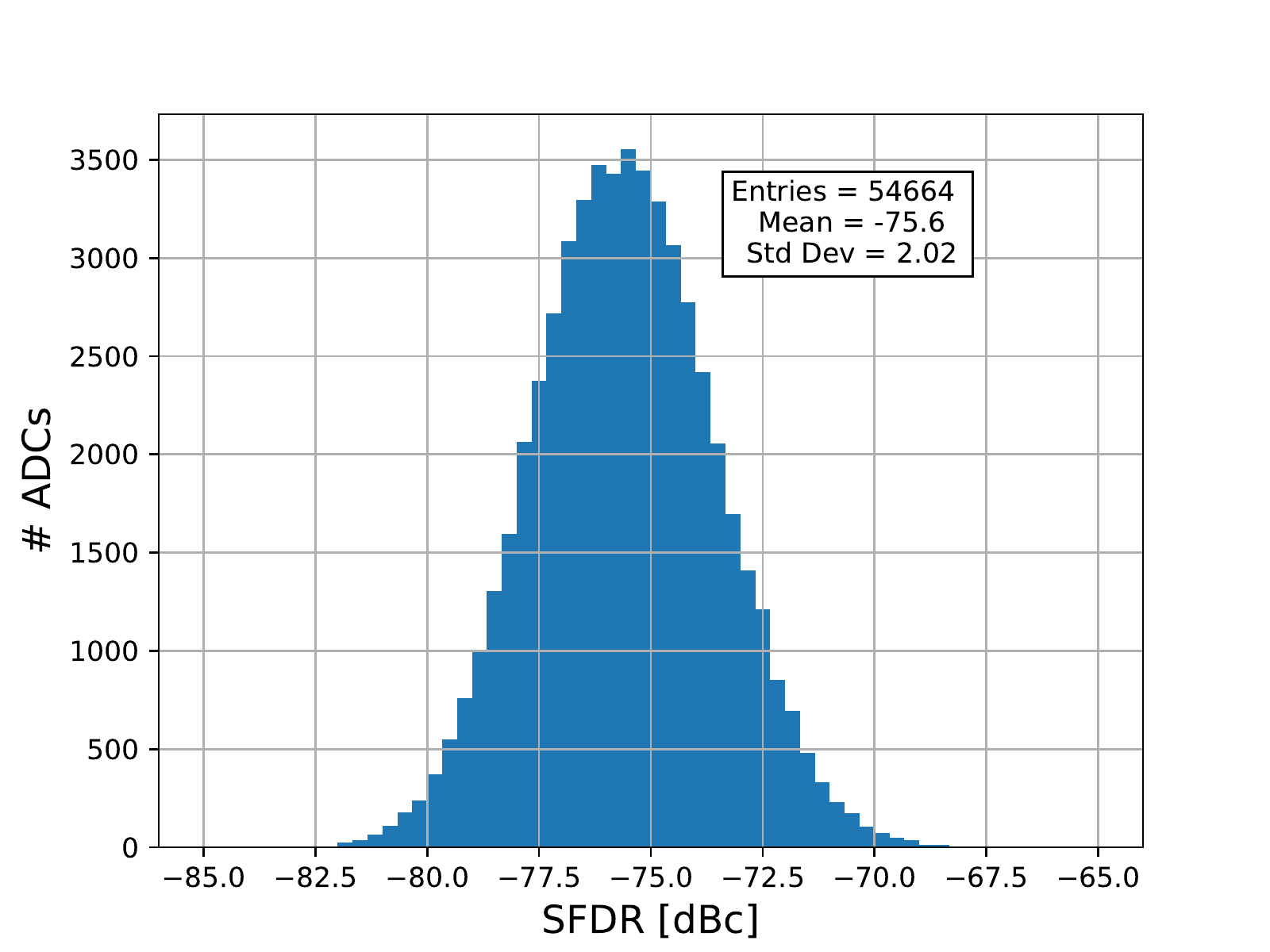}
\includegraphics[width=0.48\textwidth]{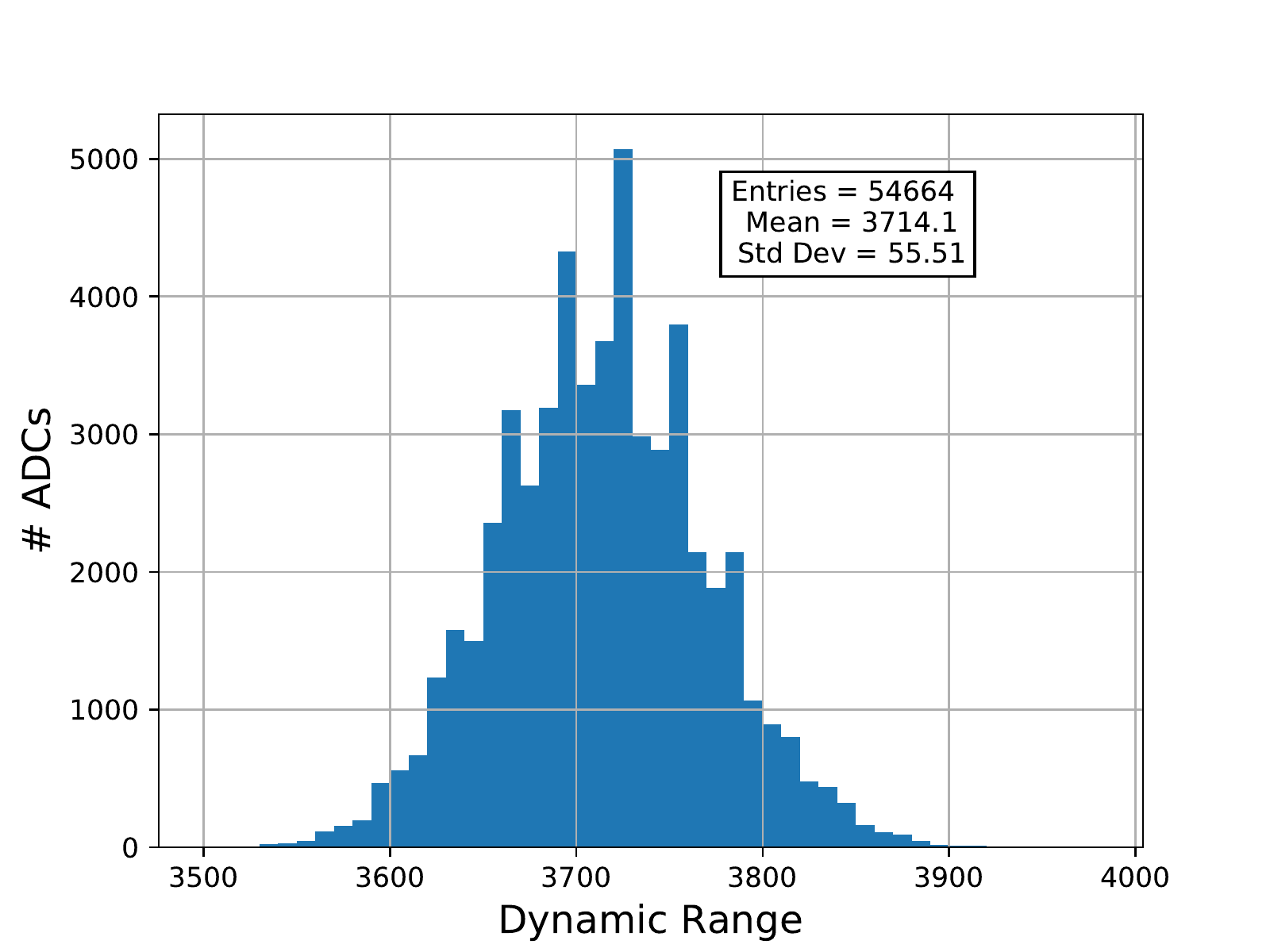}
\includegraphics[width=0.48\textwidth]{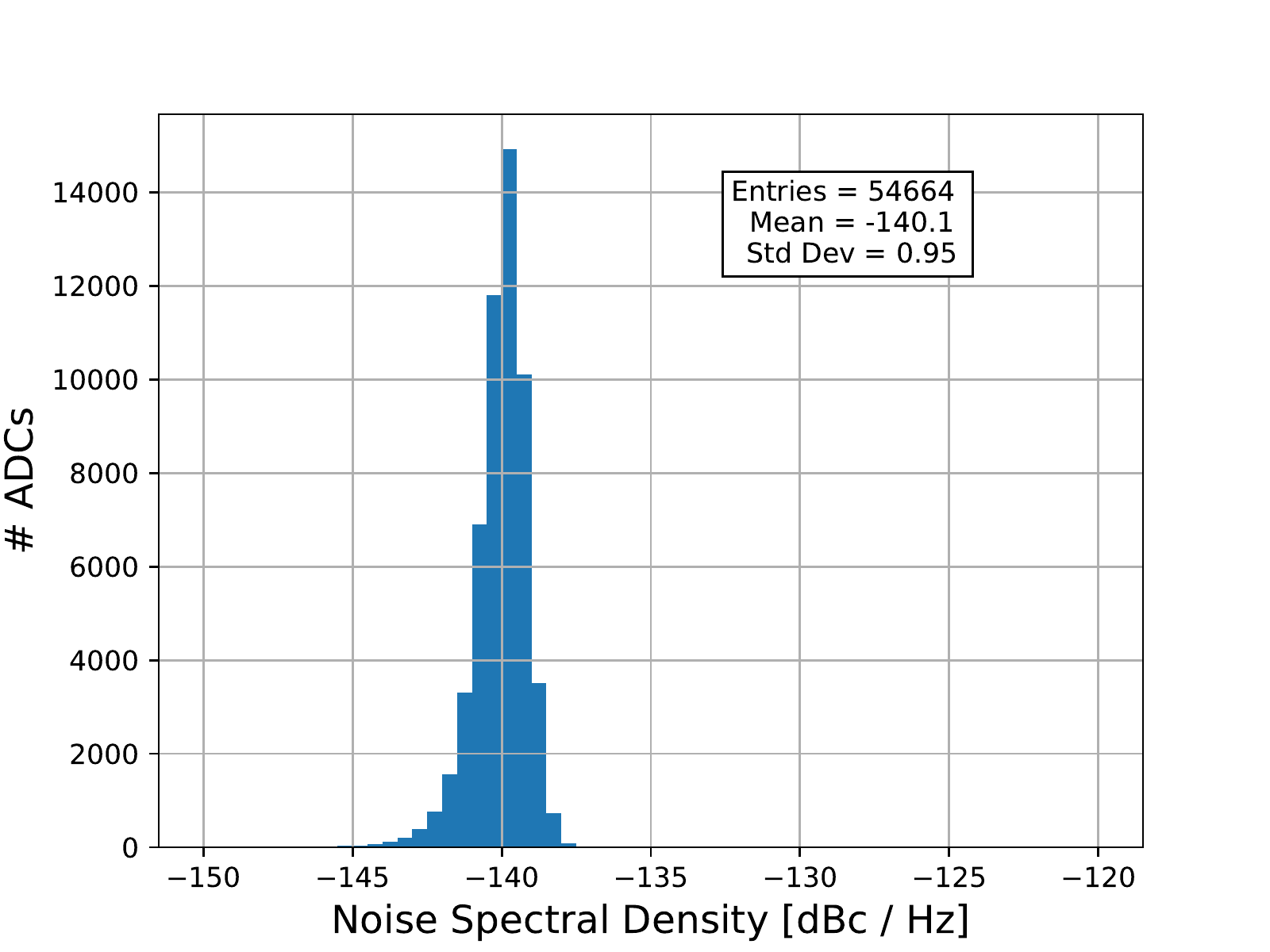}
\end{center}
\caption{Figures of merit characterizing the performance of the \totalPlotsABCD~chips in categories A, B, C and D.}
\label{fig:results}
\end{figure}

The measured FFT is used to calculate each channel's ENOB, the signal to distortion ratio (SINAD) and the spurious free dynamic range (SFDR). The corresponding distributions using the data from chips in categories A, B, C and D are shown in Figure~\ref{fig:results} and have means of 10.3-bits, 63.6~dB and 75.6~dB, respectively. The socket reduces the analog performance somewhat. Comparisons made for a few chips between measurements in sockets and with the chips soldered directly to the board indicate this reduction ranges between 0.1 and 0.5~ENOB. The dynamic range is calculated from the calibration constants and peaks at approximately 3710~ADC counts (11.9~bits), with a width of about 55 counts, or 1.5\%. Finally, the noise spectral density (NSD) is plotted, with a mean of approximately -140~dBc/Hz. 

\begin{table}[!htb]
\begin{center}
\caption{Chip categorization as a function of QC results.}
\begin{tabular}{ccc}
\hline
Category & Description & Number of chips \\
\hline
A & DR$>$3700, ENOB$>$9.9 & \totalA \\
B & DR$>$3600, ENOB$>$9.9 & \totalB \\ 
C & DR$>$3600, ENOB$>$9.5 & \totalC \\
D & DR$>$3500, ENOB$>$9.5 & \totalD \\
F & failed & \totalFailed \\ 
\hline
\end{tabular}
\label{tab:cat}
\end{center}
\end{table}

\section{Summary}

This paper summarizes the final design and quality control of a quad-channel 12-bit 40~MSPS pipeline ADC that will be installed in the ATLAS LAr calorimeter electronics readout as part of the Phase-I upgrade. The final version of the chip meets the performance requirements for operation during Run~3 and the HL-LHC and addresses the features identified during earlier design and testing stages. A total of \totalAB~chips were identified as having all four channels with dynamic range above 3600 counts and ENOB larger than 9.9, and were selected for assembly on the LTDB. 

The radiation tolerance of the ADC design was previously established for a TID up to 10~MRad, with proton SEE cross-sections measured to be of the order of 10\textsuperscript{-12}~cm\textsuperscript{2} per ADC channel. Additional measurements of SEE cross-sections were performed using beams of heavy ions, allowing for a characterization of the ADC radiation tolerance in a wider energy range with respect to previous results using protons.

\acknowledgments
We would like to acknowledge the excellent work and essential contributions of the technical staff at Nevis Laboratory, including everyone who contributed to the quality control tests: Nancy Bishop, Elena Busch, Raymond Dudley, Julia Gonski, Rupal Gupta, Alan Kahn, Kiley Kennedy, Daniel Williams. This project has received funding from the European Union's Horizon 2020 Research and Innovation programme under Grant Agreement no. 654168 and by the US National Science Foundation, Award Numbers 1707971 and 1345157.

\bibliography{paper} 

\providecommand{\href}[2]{#2}\begingroup\raggedright\begin{thebibliography}{10}

\bibitem{phase1tdr}
M.~C. Aleksa, W.~P. Cleland, Y.~T. Enari, M.~V. Fincke-Keeler, L.~C. Hervas,
  F.~B. Lanni et~al., \emph{{{ATLAS Liquid Argon calorimeter Phase-I Upgrade
  Technical Design Report}}},  Tech. Rep. CERN-LHCC-2013-017. ATLAS-TDR-022,
  CERN, Sep, 2013.

\bibitem{Evans:2008zzb}
L.~Evans and P.~Bryant, \emph{{LHC Machine}},
  \href{http://dx.doi.org/10.1088/1748-0221/3/08/S08001}{\emph{J. Instrum}
  {\bfseries 3} (2008) S08001}.

\bibitem{Collaboration_2008}
{ATLAS Collaboration}, \emph{The {ATLAS} {Experiment} at the {CERN} {Large
  Hadron Collider}},
  \href{http://dx.doi.org/10.1088/1748-0221/3/08/s08003}{\emph{J. Instrum}
  {\bfseries 3} (2008) S08003--S08003}.

\bibitem{FEB}
N.~J. Buchanan et~al., \emph{{{Design and implementation of the Front End Board
  for the readout of the ATLAS Liquid Argon calorimeters}}},
  \href{http://dx.doi.org/10.1088/1748-0221/3/03/P03004}{\emph{J. Instrum}
  {\bfseries 3} (2008) P03004}.

\bibitem{locid}
F.~Liang, D.~Gong, S.~Hou, T.~Liu, C.~Liu, D.~S. Su et~al., \emph{{{The Design
  of 8-Gbps VCSEL drivers for ATLAS Liquid Argon calorimeter Upgrade}}},
  \href{http://dx.doi.org/10.1088/1748-0221/8/01/C01031,
  10.1088/1748-0221/8/02/E02001}{\emph{J. Instrum} {\bfseries 8} (2013)
  C01031}.

\bibitem{nevis10}
J.~Kuppambatti, J.~Ban, T.~Andeen, P.~Kinget and G.~Brooijmans, \emph{{{A
  radiation-hard dual channel 4-bit pipeline for a 12-bit 40 MS/s ADC prototype
  with extended dynamic range for the ATLAS Liquid Argon calorimeter readout
  electronics upgrade at the CERN LHC}}},
  \href{http://dx.doi.org/10.1088/1748-0221/8/09/P09008}{\emph{J. Instrum}
  {\bfseries 8} (2013) P09008},
  [\href{https://arxiv.org/abs/1308.0028}{{\ttfamily 1308.0028}}].

\bibitem{nevis12}
J.~Kuppambatti, J.~Ban, T.~Andeen, R.~Brown, R.~Carbone, P.~Kinget et~al.,
  \emph{{{A radiation-hard dual-channel 12-bit 40 MS/s ADC prototype for the
  ATLAS Liquid Argon calorimeter readout electronics upgrade at the CERN
  LHC}}}, \href{http://dx.doi.org/10.1016/j.nima.2017.01.025}{\emph{Nucl.
  Instrum. Meth.} {\bfseries A855} (2017) 38--46},
  [\href{https://arxiv.org/abs/1706.01535}{{\ttfamily 1706.01535}}].

\bibitem{radiation_qualification}
N.~J. Buchanan et~al., \emph{{{Radiation qualification of the front-end
  electronics for the readout of the ATLAS Liquid Argon calorimeters}}},
  \href{http://dx.doi.org/10.1088/1748-0221/3/10/P10005}{\emph{J. Instrum}
  {\bfseries 3} (2008) P10005}.

\bibitem{phase2tdr}
{ATLAS Collaboration}, \emph{{{Technical Design Report for the Phase-II Upgrade
  of the ATLAS LAr calorimeter}}},  Tech. Rep. CERN-LHCC-2017-018.
  ATLAS-TDR-027, CERN, Geneva, Sep, 2017.

\bibitem{WeibullPaper}
L.~D.~Edmonds, \emph{{{Proton SEU Cross Sections Derived from Heavy-Ion Test
  Data}}}, \href{http://dx.doi.org/10.1109/TNS.1977.4328890}{\emph{IEEE
  Transactions on Nuclear Science} {\bfseries 47} (2000) 1713--1728}.

\bibitem{dEdx}
M.~Berger, J.~Coursey, M.~Zucker and J.~Chang, \emph{{NIST Standard Reference
  Database 124: Stopping-Power \& Range Tables for Electrons, Protons, and
  Helium Ions - Version 2.0.1, National Institute of Standards and
  Technology}},  2017.
\newblock https://dx.doi.org/10.18434/T4NC7P.

\end{thebibliography}\endgroup
\bibliographystyle{JHEP}
\end{document}